\documentclass[12pt,preprint]{aastex}

\newcommand{\eb}{\begin{equation}}
\newcommand{\ee}{\end{equation}}

\newcommand{\kms}   {\ensuremath{ \mbox{km\,s}^{-1}                 }}

\newcommand{\kmspc} {\ensuremath{ \mbox{km\,s}^{-1}\,\mbox{kpc}^{-1} }}

\newcommand{\masyr} {\ensuremath{ \mbox{ mas\,yr}^{-1} }}

\shorttitle{Stellar velocity field}
\shortauthors{Makarov \& Murphy}

\begin{document}

\title{The local stellar velocity field via vector spherical harmonics}

\author{V. V. Makarov}
\affil{Michelson Science Center, California Technology Institute, 770 S. Wilson Ave.,
MS 100-22, Pasadena, CA 91125}
\email{vvm@caltech.edu}
\and
\author{D. W. Murphy}
\affil{Jet Propulsion Laboratory, California Technology Institute, 4800 Oak Grove Drive,
MS  138-308, Pasadena, CA 91105}
\email{david.w.murphy@jpl.nasa.gov}

\begin{abstract}

We analyze the local field of stellar tangential velocities for a
sample of $42\,339$ non-binary Hipparcos stars with accurate
parallaxes, using a vector spherical harmonic formalism. We derive
simple relations between the parameters of the classical linear model
(Ogorodnikov-Milne) of the local systemic field and low-degree terms
of the general vector harmonic decomposition. Taking advantage of
these relationships we determine the solar velocity with respect to
the local stars of $(V_X,V_Y,V_Z)=(10.5,\, 18.5,\, 7.3)\pm 0.1$ \kms,
not corrected for the asymmetric drift with respect to the Local
Standard of Rest (LSR). If only stars more distant than 100 pc are
considered, the peculiar solar motion is $(V_X,V_Y,V_Z)=(9.9,\,
15.6,\, 6.9)\pm 0.2$ \kms. The
adverse effects of harmonic leakage, which occurs between the reflex
solar motion represented by the three electric vector harmonics in the
velocity space and higher-degree harmonics in the proper motion space,
are eliminated in our analysis by direct subtraction of the reflex
solar velocity in its tangential components for each star. The Oort's
parameters determined by a straightforward least-squares adjustment in
vector spherical harmonics, are $A=14.0\pm 1.4$, $B=-13.1\pm 1.2$,
$K=1.1\pm 1.8$, and $C=-2.9\pm 1.4$ \kmspc.  The physical meaning and
the implications of these parameters are discussed in the framework of
a general linear model of the velocity field. We find a few
statistically significant higher degree harmonic terms, which do not
correspond to any parameters in the classical linear model. One of
them, a third-degree electric harmonic, is tentatively explained as
the response to a negative linear gradient of rotation velocity with
distance from the Galactic plane, which we estimate at $\sim -20$
\kmspc.  A similar vertical gradient of rotation velocity has been
detected for more distant stars representing the thick disk ($z > 1$
kpc), but here we surmise its existence in the thin disk at $z < 200$
pc. The most unexpected and unexplained term within the
Ogorodnikov-Milne model is the first-degree magnetic harmonic
representing a rigid rotation of the stellar field about the axis $-Y$
pointing opposite to the direction of rotation. This harmonic comes
out with a statistically robust coefficient $6.2 \pm 0.9$ \kmspc, and
is also present in the velocity field of more distant stars. The
ensuing upward vertical motion of stars in the general direction of
the Galactic center and the downward motion in the anticenter
direction are opposite to the vector field expected from the stationary Galactic
warp model.
\end{abstract}


\section{Introduction}
The growing amount and quality of available astrometric data and the
widening horizons of the astrometrically known Galaxy begin to call
for more general and accurate models of the Galactic velocity field
than the classical Oort and Lindblad approximation. The classical
Oort's constants $A$ and $B$ representing the local angular velocity
of rotation and its radial gradient have been determined many times on
different data sets with increasing accuracy, but only the advent of
the Hipparcos catalog made it possible to put this estimation on a
systematically rigid, global reference frame related to a non-rotating
extragalactic frame.  A somewhat heuristic generalization of the
Oort's formalism, known as the Ogorodnikov-Milne model (OMM
hereafter), describes the local field of stellar velocities as a
combination of all six possible shears and three rigid rotations,
incorporating the $A$ and $B$ parameters, and adding seven more
degrees of freedom to the fitting model.  Not all of these additional
parameters have a clear physical meaning in terms of Galactic
dynamics. To our knowledge, only two additional coefficients ($K$ and
$C$) have been paid close attention to in the recent literature. The
remaining five OMM parameters have been tacitly considered
non-essential and presumably small. Furthermore, as a linear
approximation to a complex and, on a large scale, nonlinear Galactic
velocity field, the Oort's formalism and the OMM are meaningful only
locally, within reasonably small distances from the Sun. As we expand
the limits of precision astrometric catalogs, we inevitably encounter
significant deviations from the linear model. Following the suggestion
of \citet{olme}, one could consider the OMM parameters to be functions
of heliocentric distance, or, in a more general approach, of galactic
cylindrical coordinates. Alternatively, as we investigate in this
paper, one can use a more versatile formalism for describing the
observed velocity field in order to put this problem on a more
rigorous and systematic footing.

The paper by \citet{vit} marked an important advance in the search for
a better way to describe the systematic field of tangential velocities
derived from Hipparcos proper motions and parallaxes for tens of
thousands stars.  They employed vector spherical harmonic functions,
which constitute an orthogonal basis of the space of continuous vector
functions on unit sphere. When sampled over a large and sufficiently
uniform set of points on the celestial sphere, the discretized vector
harmonic functions remain nearly orthogonal, which makes the fitting
algorithm the most stable and accurate in implementation.  Since the
Oort's and OMM parameters turn out to be simply the fitting
coefficients of the corresponding low-degree vector harmonic
functions, the physical meaning and interpretation of the estimated
quantities remain straightforward and simple.  The mathematical
formulation allows us not only to accurately estimate the
uncertainties associated with the estimated parameters (arising mostly
from the stochastic component of the velocity field), but also, for
the first time, to investigate the covariances. But the main advantage
of this method is that it is easily expandable to handle more complex
and nonlinear fields. The aims of this paper are: 1) using the vector
harmonic method, to re-determine the fundamental parameters of the
local velocity field in a rigorous fashion, viz., the peculiar solar
velocity and the differential rotation coefficients $A$ and $B$; 2) to
estimate other OMM parameters; 3) to find out if the observed velocity
field bears statistically significant, higher-degree non-OMM harmonic
terms that may reveal interesting dynamical phenomena, such as
vertical rotation gradients, Galactic warp and large streams of stars.

\section{The peculiar velocity of the Sun}
\label{sun.sec}
\begin{figure}
\plotone{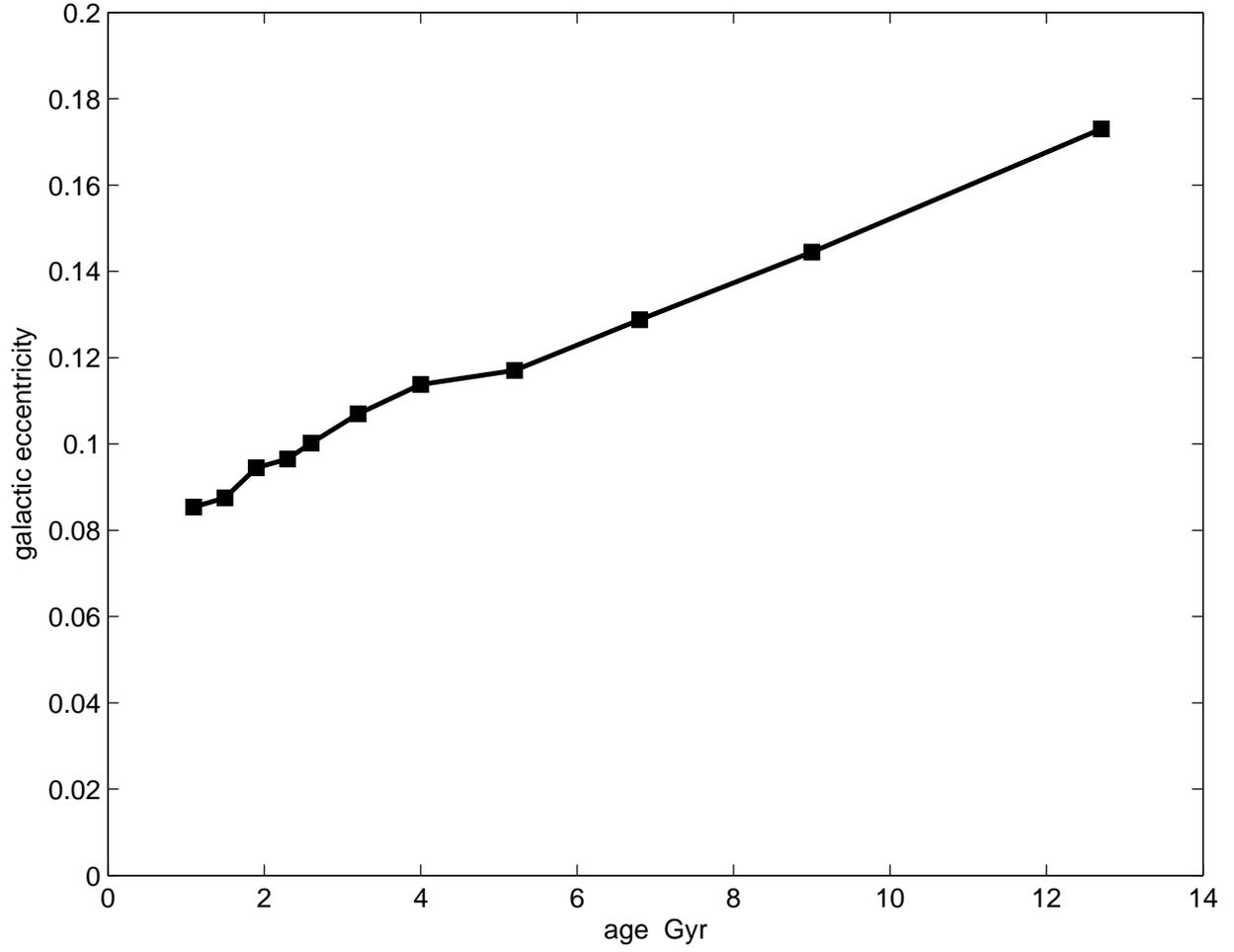}
\caption{Dependence of mean orbital eccentricity on stellar age. Data extracted from the Geneva-Copenhagen spectroscopic survey of Hipparcos stars
\citep{nord}. \label{ecc.fig}}
\end{figure}

The dynamic LSR is defined as a fictitious point currently at the
position of the Sun in the Galactic plane, which moves along a
perfectly circular orbit in a hypothetical axisymmetric
potential. This definition is a mere theoretical concept, because the
Galactic potential is not exactly axisymmetric, and there are no
perfectly circular orbits. An alternative {\it empirical} definition
of the LSR is the average motion of a sufficiently large, sufficiently
homogeneous and sufficiently dynamically mixed sample of stars
centered on the Sun. These two definitions are quite different, and in
fact, are contradictory with regard to the shape of the reference
orbit.  Indeed, the empirical
average reference orbit is markedly eccentric due to the stellar age-dependent asymmetric
drift (see Fig.~\ref{ecc.fig}). The relative velocity of the Sun with respect to the empirical
LSR is directly derived from our heliocentric astrometric
observations, whereas determination of the solar {\it peculiar}
velocity with respect to the hypothetical circular orbit is more
involved \citep{mih, deh} and relies on additional astrophysical or
dynamical considerations.

If the Sun moves with a velocity $\vec{V}_{\sun} = (V_X, V_Y, V_Z)$
relatively to the average motion of the local stars, the heliocentric
velocity field of these stars contains a streaming motion in the
opposite direction, that is, $-\vec{V}_{\sun}$. As far as tangential
velocity components are concerned, this streaming motion is a dipolar
vector field on the celestial sphere, and its exact representation via
vector spherical harmonics (see Eqs.~\ref{ele.eq}) is

\eb
\vec{v}_\tau = V_X \vec{E}_1^1 + V_Y \vec{E}_1^{-1} - V_Z \vec{E}_1^0.
\ee

Thus, the signature of the solar motion is confined to the first three
electric harmonics only. We will see in the subsequent paragraphs that
the other fundamental parameters of the velocity field are represented
by magnetic harmonics and electric harmonics of higher degree, so that
this effect is clearly separated in the vector harmonic space. It is
worth noting that $\vec{V}_{\sun}$ is estimated from tangential
velocities in physical units, i.e., from $\mu /\Pi$, where $\mu$ is
the proper motion magnitude, and $\Pi$ is the parallax. A small
admixture of halo stars and runaway stars with very high spatial
velocities can perturb this determination.

We start with selecting $42\,487$ stars from the main Hipparcos
catalog with statistically robust parallaxes ($\Pi/ \sigma(\Pi) > 5$)
and without any indicators of binarity. To avoid extra-high velocity
perturbers, we reject 148 stars with tangential velocities greater
than 150 \kms~ in either galactic component. A set of 24 vector
harmonic functions is then fitted to the global vector field of
$42\,339$ stars by a direct least-squares solution. The solar velocity
components are simply read from the fitted coefficients of the first
three electric harmonics. The results of this estimation are specified
in Table~\ref{vsun.tab}, along with a sample of more distant stars
with $\Pi > 10$ mas ($24\, 327$ stars).

The estimated velocity components $V_X$ and $V_Z$ are fairly similar
for the two samples, indicating a negligible dependence on
distance. They are also very close to the determination by
\citet{deh}. The estimates of $V_Y$ (in the direction of Galactic
rotation) are very different between the two samples, beyond the
possibility of a statistical fluke. The more distant stars move faster
with respect to the Sun than the stars closer in. In either case, the
stars move slower in this direction by more than 10 \kms~ than the
circular motion of the LSR determined by \citet{deh}.  This very
prominent effect is attributed to the asymmetric drift of nearby
stars, discussed in more detail in the next paragraph. Generally,
there are three different reasons of physical and technical kind for
our estimation of $V_Y$ to be biased:

\begin{enumerate}
\item
the asymmetric drift;
\item
the vertical gradient of rotational velocity $\Gamma(z)$;
\item
the mixing of non-orthogonal harmonics.
\end{enumerate}

\section{The physical meaning of Oort's constants}

\label{oo.sec}

As long as a relatively small local area of the Galaxy around the Sun
is considered ($r << R_0$, where $R_0$ is the galactocentric distance
of the Sun), it is appropriate to expand the systemic velocity field
of stars in a Taylor series over the galactocentric cylindrical
coordinates $(\rho, \theta, z)$. Assuming that the local field is
planar, that is, all systemic motions are in the galactic plane, the
velocity vector $\vec{v}(\rho, \theta, z)=\vec{a}(\rho,
\theta)+\vec{\omega}(\rho, \theta, z)$, where the former component
$\vec{a}$ is radial with respect to the Galactic center, and the
latter, $\vec{\omega}$ is tangential and orthogonal to the former.
Note that the radial component is assumed to be independent of $z$,
that is, that there is no radial systemic motion depending on the
distance from the plane. Retaining only first-degree terms in the
corresponding Taylor expansion, one can write

\begin{eqnarray}\label{taylor.eq}
a(\rho, \theta)&=&a_0+\delta\,(\rho-\rho_0)+\epsilon\, \sin (\theta - \theta_0)\\\nonumber 
\omega(\rho, \theta, z)&=&\omega_0+\alpha\, (\rho-\rho_0)+\beta\, \sin (\theta - \theta_0)+\Gamma(z)\nonumber 
\end{eqnarray}

where the subscript 0 denotes the corresponding parameters at the
Sun's location $(\rho_0, \theta_0, z_0)$.  We left the dependence of
the rotational velocity $\omega$ on $z$ in its generic form, since on
physical grounds, this dependence is expected to be symmetric around
the plane, and can not be represented by a simple linear term.

Retaining only terms to $O(\frac{r}{\rho_0})$, these model relations
can be rewritten more conveniently in the heliocentric coordinates
$(r, \ell, b)$ introduced in Appendix A:

 \begin{eqnarray}\label{model.eq}
a(\vec r)&=&a_0-\delta\,r\cos \ell\, \cos b+\epsilon\, \frac{r}{\rho_0}\sin \ell \,\cos b \\\nonumber 
\omega(\vec r)&=&\omega_0-\alpha\, r\, \cos \ell \, \cos b+\beta\, \frac{r}{\rho_0}\sin \ell \,\cos b+\Gamma(z)
\nonumber \end{eqnarray}

Apart from the reflex peculiar motion of the Sun treated in Section
\ref{sun.sec}, we observe the heliocentric velocity field $\Delta \vec
v = \vec a + \vec \omega - \vec a_0 -\vec \omega_0$. Projections of
this vector field onto the local tangential coordinate frames $(\vec
\tau_\ell, \vec \tau_b)$ introduced in Appendix A, are

 \begin{eqnarray}\label{proj.eq}
\Delta \vec v \cdot \vec \tau_\ell&=&\frac{\omega(\vec r)}{2\rho_0}\, r\, \cos 2\ell\, \cos b - \frac{\omega(\vec r)}{2\rho_0}\, r\,\cos b
+(\omega(\vec r)-\omega_0)\, \cos \ell \\\nonumber 
& & + \frac{a(\vec r)}{2\rho_0}\, r\, \sin 2\ell\, \cos b + (a(\vec r)-a_0)\, \sin \ell \\\nonumber
\Delta \vec v \cdot \vec \tau_b  &=&-\frac{\omega(\vec r)}{2\rho_0}\, r\, \sin 2\ell\, \cos b\, \sin b- 
(\omega(\vec r)-\omega_0)\, \sin \ell \, \sin b \\ 
& & - \frac{a(\vec r)}{2\rho_0}\, r\, \cos b \, \sin b + \frac{a(\vec r)}{2\rho_0}\, r\, \cos 2\ell\,\cos b \, \sin b+(a(\vec r)-a_0)\, \cos \ell \, \sin b
\nonumber \end{eqnarray}

Substituting the model (\ref{model.eq}) into Eqs.~\ref{proj.eq} and
retaining only terms to $O(\frac{r}{\rho_0})$, we obtain after some
toil the general expansion

 \begin{eqnarray}\label{exp.eq}
\Delta \vec v &=&\left(\frac{\omega_0}{2\rho_0}-\frac{\alpha}{2}-\frac{\epsilon}{2\rho_0}\right)\, r\, \frac{1}{6}\, \vec E_2^{-2}
 + \left(-\frac{\omega_0}{2\rho_0}-\frac{\alpha}{2}+\frac{\epsilon}{2\rho_0}\right)\, r\, \vec H_1^{0}    \nonumber \\
& & + \left(\frac{a_0}{2\rho_0}-\frac{\delta}{2}-\frac{\beta}{2\rho_0}\right)\, r\, \frac{1}{6}\, \vec E_2^{2}
 + \left(-\frac{a_0}{2\rho_0}-\frac{\delta}{2}-\frac{\beta}{2\rho_0}\right)\, r\, \frac{1}{3}\,\vec E_2^{0}    \nonumber \\
& & +\frac{\Gamma(z)}{2\rho_0}\, r\, \frac{1}{6}\, \vec E_2^{-2} - \frac{\Gamma(z)}{2\rho_0}\, r\, \vec H_1^{0}-\Gamma(z)\,
\vec E_1^{-1}
\end{eqnarray}
where we made use of the functional forms of the vector spherical
harmonics in galactic coordinates specified in Appendix A.
Disregarding for now the $\Gamma(z)$ terms, let us compare this
equation with the classical expansion of the velocity field via the
fundamental Oort's constants \citep[e.g.,][]{tor}, which in the vector
harmonics notation takes the form

\eb
4.741\, r\,\vec \mu=A\cdot \frac{1}{6}\,\vec E_2^{-2}+B\cdot \vec H_1^0+C\cdot \frac{1}{6}\,\vec E_2^{2}-K\cdot \frac{1}{3}\,\vec E_2^{0}.
\ee
Hence, in our more general model, the Oort's constants can be defined as
 \begin{eqnarray}\label{oort.eq}
A & = & \frac{\omega_0}{2\rho_0}-\frac{\alpha}{2}-\frac{\epsilon}{2\rho_0}  \\
B & = & -\frac{\omega_0}{2\rho_0}-\frac{\alpha}{2}+\frac{\epsilon}{2\rho_0} \\
C & = & \frac{a_0}{2\rho_0}-\frac{\delta}{2}-\frac{\beta}{2\rho_0} \\
K & = & \frac{a_0}{2\rho_0}+\frac{\delta}{2}+\frac{\beta}{2\rho_0}.
\end{eqnarray}

The slope of the local rotational velocity curve is readily derived as
$\alpha = -(A+B)$. Since most of the recent and Hipparcos-based
estimations arrive at $A\approx -B$, the rotation curve is locally
almost (but not exactly) flat. It is also usually adopted that the
local angular rotation velocity $\dot \theta_0$ is just the difference
of the constants $A$ and $B$. In fact, however,

\eb
\label{theta.eq}
\dot \theta_0 = \frac{\omega_0}{\rho_0} = A-B+\frac{\epsilon}{\rho_0}.
\ee

So, the local azimuthal shear of the radial motion $\epsilon$
contributes to the constants $A$ and $B$ and affect the determination
of the angular velocity of the Galaxy. Presumably, this shear is small
($\epsilon << \omega_0$), and our estimations are not hampered too
much. The interpretation of the constants $C$ and $K$ is more
complicated. Traditionally, the $C$ constant is called the shear, and
the $K$ constant the local expansion (or dilation). In fact, we find
that

\begin{eqnarray}
C +K& = & \frac{a_0}{\rho_0} \label{c_k.eq}\\
K-C  & = & \delta +\frac{\beta}{\rho_0}. \label{k_c.eq}
\end{eqnarray}

Presumably, there is no outward or inward bulk motion of stars around
the Sun ($a_0=0$), and the LSR, empirically defined as the mean motion
of a large homogenous heliocentric sample of stars, moves on a
circular orbit. In numerous and somewhat conflicting determinations,
it appears that, generally, $C\neq -K$, which indicates a nonzero
systemic eccentricity of the local field. It is especially important
in this case to accurately estimate the uncertainties of estimation,
which may be larger than the estimates, as discussed in Section
\ref{fit.sec}. The dilation is better characterized by the difference
$K-C$ in Eq.~\ref{k_c.eq}, where $\delta$ can be interpreted as the
radial heliocentric expansion, and $\beta/\rho_0$ as the azimuthal
expansion. These two components can not be separated from proper
motions alone.

\section{Parameters of the velocity field}
\label{fit.sec}

Using the vector harmonic formalism to describe the tangential
velocity field on the celestial sphere (Appendix A) and the general
expression for Ogorodnikov-Milne model (Eq.~\ref{om.eq}), makes the
estimation of model parameters quite straightforward. The Hipparcos
proper motions for our initial set of $42\,339$ non-binary stars with
accurate parallaxes are converted to tangential velocities in the
galactic coordinate system. Each velocity vector generates two
condition equations, one for the longitudinal component
$\vec{\tau}_{\ell}$ and the other for the latitudinal component
$\vec{\tau}_b$. The coefficients of the expansion \ref{exp.eq} are the
unknowns of the condition equations, which are solved by the
least-squares method. The main source of the solution uncertainty is
the physical dispersion of individual velocity vectors related to
peculiar orbital motions, since the astrometric errors of proper
motions are small in the Hipparcos catalog, and the number of stars is
large. The velocity dispersion is known to increase with age; it is
also larger for thick disk stars than for thin disk stars. Instead of
dealing with triaxial dispersion ellipsoids for various stellar
populations, we take an empirical and robust approach to error
estimation in this least-squares adjustment.  We fit 24 vector
harmonic functions up to degree 4 to the general vector field and
consider this expansion to represent the systemic part of the velocity
field. The residuals of the tangential velocity vectors represent the
stochastic part of the field. Dispersions of velocities are computed
from these residuals, separately for the $\vec{\tau}_{\ell}$ and
$\vec{\tau}_b$ components, as half-differences between the 0.84 and
0.16 quantiles on each distribution. These quantities substitute
standard deviation parameters for the markedly non-Gaussian velocity
distributions. The resulting dispersions are $\sigma(v_\ell) = 24.0$
\kms, $\sigma(v_b) = 16.1$ \kms. The condition equations in longitude
and in latitude are weighted with these quantities, respectively.

There are a few important technical notes to be made on this
estimation problem. The solar peculiar velocity vector is determined
directly from the tangential velocity field, expressed in units of
\kms~ (\S \ref{sun.sec}), in which case only the first three electric
harmonics are of essence. The Oort or Ogorodnikov-Milne parameters
describe a velocity field which grows linearly with distance from the
Sun, and has distance $r$ in its functional form (\ref{v.eq}).  The
corresponding decomposition is done in the proper motion field, or, as
we do it in this paper, the tangential velocities can be used in the
observational part of the equations, but the harmonic functions are
pre-multiplied with distances for each star.  In the latter case, the
distribution of sample stars on distance is taken into account
automatically, and the harmonic coefficients have the desired
dimension of \kms~ kpc$^{-1}$. But before performing this
distance-weighted least-squares estimation, the relative velocity of
the Sun with respect to the stellar centroid should be subtracted for
each star. This step proves to be of crucial importance because of the
adverse effects of the harmonic leakage, discussed in
\S~\ref{mix.sec}.

The results of vector harmonic estimation are specified in Table \ref
{om.tab} for the original set of $42\,339$ stars, and for $24\, 327$
stars more distant than 100 pc. All harmonic coefficients
corresponding to Ogorodnikov-Milne parameters are shown, as well as
other statistically significant coefficients which have no
counterparts in the linear model. By statistically significant we
conservatively mean a quantity larger than its formal error multiplied
by $2.5$. Having the significance so defined we state that only three
or four model OMM parameters are significant, and three extra
non-linear parameters. The estimated parameters agree very well
between the two sets, indicating no considerable dependence of the
velocity field on distance within this fairly small volume. The slope
of the rotation curve, using the results for the larger sample, is
$\alpha = -(A+B) = -1.0\pm 1.8$ \kms~ kpc$^{-1}$. This implies that
the speed of rotation declines very slightly with galactocentric
distance, but the conclusion is not reliable statistically. From
Eq.~\ref{theta.eq}, ignoring the possible contamination by the
azimuthal shear $\epsilon$, the local angular rotation is $A-B=
27.1\pm 1.8$ \kms~ kpc$^{-1}$, in fine agreement with
\citet{fea}. Assuming a distance $\rho_0=7.9$ kpc for the Sun
\citep[see][and references therein]{val}, the speed of rotation is
$\omega_0 = 214\pm 14$ \kms.

The $K$ constant is insignificant for both samples, but the $C$
constant is marginally significant, especially for more distant
stars. The systemic outward motion $a_0= \rho_0(C+K)$ appears to be
negligible for the general sample, but the more distant stars seem to
exhibit an inward motion of $a_0= -42 \pm 20$ \kms. This result is
qualitatively consistent with the estimation by \citet{han87} who
found a progressively smaller solar velocity toward the Galactic
center with respect to stars at higher latitudes. If this inward
motion in the outer part of the astrometrically known Galaxy is real,
it may be somewhat counterbalanced by the small but persistent
dilation (expansion) of the local stellar aggregate, at $K-C\simeq 4
\pm 2$ \kms~ kpc$^{-1}$. Associations of young stars expand by virtue
of their initial velocity dispersions \citep{makol}, and the presence
of the young Local Association could be invoked to explain the local
expansion. It is worth emphasizing that the accuracy of the available
astrometric data on the local stellar field is still insufficient to
establish these subtle effects with certainty. In fact, the barely
noticeable $K$ and $C$ constants may be related to the
intermediate-scale streams of stars permeating the solar neighborhood,
rather than to the general pattern of Galactic rotation. \citet{fam}
present a scrutiny of such streams or superclusters, based on the best
available radial velocity and astrometry data for K and M giants,
including the Hyades, Sirius, Hercules and B streams. Apart from
strong evidence for asymmetric drift for evolved stars, they find,
interestingly, that the centroid velocity of the Sun $V_X$ with
respect to giants is $\simeq 10$ \kms, in agreement with our present
and other previous estimations, but it drops to only $\simeq 3$ \kms~
when all the major streams are excluded. This difference may be
interpreted as a net outward radial motion of the streams (see also
their Table 1). \citet{fam} point out that the members of these
streams have a spread of ages and other physical characteristics, and
the streams must be dynamically induced. The authors raise the
question of how the standard solar motion can actually be defined if the motion
of even the oldest and supposedly dynamically mixed stellar
populations is subject to unknown dynamical agents perturbing their
orbits? We think that the stellar streams are legitimate parts of the
local velocity field, and that it makes sense to define the velocity
centroid and the solar peculiar motion in much the same way as it has
been done before, keeping in mind that dynamical mixing and relaxation
may be a mere theoretical idealization, as well as a circularly moving
LSR.

\subsection{Harmonic leakage}
\label{mix.sec}

As specified in Table \ref {om.tab}, we find only three significant
terms in the general vector harmonic decomposition beyond the
Ogorodnikov-Milne model, viz., $H_2^{-1}$, $E_3^{-1}$, and
$E_4^{2}$. All other estimated harmonics, including all third and
fourth degree terms, are well below $2.5\sigma$. Thus, we find little
evidence of nonlinear patterns in the motion of local stars. The
actual velocity field progressively deviates from the linear
approximation of the model with heliocentric distance. Furthermore,
the rotation curve may have local wiggles and curvature, as discussed
in \citet{olme}. One way of tackling this problem is to build a more
complex model in which the Oort's constants are actually functions of
coordinates, to be determined from observations. We take a different
approach in this paper, determining empirically a vector harmonic
decomposition and trying to interpret those terms that appear to be
statistically significant.

Before embarking on analysis of the emerging nonlinear harmonics (and
the unexpected linear term $H_1^{-1}$), we should examine a technical,
but crucial problem in the determination of model parameters. The
stellar velocity field bears a strong signal in the classical terms
representing the reflex solar motion and the Oort's constants $A$ and
$B$. These terms are represented in our model by specific vector
harmonics (Appendix B). The strong signal in the physically meaningful
low-degree harmonics in the velocity space can leak into higher-degree
vector harmonics in the proper motion space, resulting in spurious
detections of nonlinear effects. This inevitably happens because the
sampled vector harmonics are not independent for any inhomogeneous
discrete set of points. This problem has two somewhat different
aspects. In classical applications, when only proper motions are known
from observation, the mean parallax of nearby stars varies across the
sky because of the real clumps in number density (the Gould Belt,
large associations, spiral arms), as well as the non-uniform
interstellar extinction.  This difficulty was first spotted by
\citet{edm}, and later investigated in more detail by \citet{olde}. In
the latter paper, a nice example is presented, how the longitudinal
variation of the mean parallax, described by a Fourier series, makes
the simple dipolar pattern of the solar motion to contribute to the
terms that would be empirically defined as the $A$ and $B$
constants. Our analysis is free of this complication, because we use
accurate trigonometric parallaxes from the Hipparcos catalog, and
perform the estimations of the solar motion in the velocity space, and
of model parameters separately in the proper motion space. But there
is another, more basic reason to be concerned about the harmonic
leakage. The lack of uniformity in the number density of stars on the
sky itself makes the vector harmonics mutually dependent within either
coordinate component.

Mathematically, the problem can be viewed as a lack of orthogonality
of the sampled harmonics. The degree of non-orthogonality is
quantified by the correlation coefficients readily computed from the
off-diagonal elements of the covariance matrix.  For our nearby stars,
the largest physical effect is the solar motion expressed by the
first-degree electric harmonics, and the cross-talk of these terms
with other harmonics of higher degree may generate false positive
detections.

We set up a dedicated numerical experiment to prove that this
contamination may happen unless appropriate precautions are taken.  We
use the same general set of Hipparcos stars as before, but the actual
observed proper motions are replaced with simulated vectors, computed
from the reflex solar motion only, estimated in \S \ref{sun.sec}. A
harmonic decomposition of the simulated velocity field produces the
same velocity dipole in the first electric harmonics, and zero for the
rest of harmonics, which only shows that the software works
correctly. But when a similar decomposition is carried out in the
space of distance-weighted harmonics, as described in \S
\ref{fit.sec}, a number of spurious terms emerge, viz.,
$\vec{H}_2^{-1}$ (significance level $8.8\sigma$), $\vec{H}_2^{0}$
($5.4\sigma$), $\vec{H}_2^{2}$ ($3.2\sigma$), $\vec{E}_3^{-3}$
($8.9\sigma$), $\vec{E}_3^{-1}$ ($4.0\sigma$), $\vec{E}_3^{0}$
($7.6\sigma$), $\vec{E}_3^{2}$ ($8.7\sigma$), and $\vec{E}_3^{3}$
($3.8\sigma$). The appearance of the $\vec{H}_2^{-1}$ and
$\vec{E}_3^{-1}$ harmonics is especially worrisome, because they may
carry some physical information, as discussed in subsequent
paragraphs. The simplest way to get rid of most of the harmonic
leakage effect is to subtract the reflex solar motion from all
tangential velocities prior to a model parameter fitting. Ideally this
eliminates the perturbations from the dominating dipolar terms. Note
that existing correlations between the sampled distance-weighted
vector harmonics that we use to determine OMM parameters, do not
affect the results in a systematic way, because the least-squares
solution is unbiased. The major adverse effect of these correlations
is an enhanced propagation of random and possibly systematic errors
from our observational data.

\section{Vertical gradient of rotational velocity}
\label{gamma.sec}

The apparent relative velocity of the Sun in the direction of galactic
 rotation ($V_{Y \sun}$) varies with distance of reference field stars
 from the Galactic plane. This remarkable fact was established by
 \citet{han89} from proper motions of fairly distant stars, and
 recently confirmed by \citet{gir}, who used absolute proper motions
 of giant stars in the direction of south Galactic pole. The thick
 disk dominates between $z=1$ and $3$ kpc, where the rotational lag of
 field stars is found to follow a nearly linear dependence on vertical
 height, accompanied, predictably, by a growth of velocity dispersion
 in the $X$ direction. The slope of the lag, from both cited papers,
 is estimated at $-30$ \kmspc. Girard et al. also offer a dynamical
 interpretation of this phenomenon, finding it consistent with a
 general model of the Galactic potential. The sample of nearby
 Hipparcos stars considered in this paper is practically limited to
 200 pc, and is dominated by thin disk stars. Is there a similar
 vertical gradient of rotational velocity for the thin disk?

Evidently from Eq.~\ref{exp.eq}, a vertical lag affects the
determination of the centroid velocity $V_Y$ expressed by the dipole
vector harmonic $\vec{E}_1^{-1}$, because $\Gamma(z)$ is negative
everywhere except $z=0$. If the velocity of rotation falls off with
increasing $z$, the relative solar velocity $V_{Y \sun}$ should grow
with distance due to the admixture of high-$z$ stars. This is not what
we find in Table~\ref{vsun.tab}, where the more distant stars ($\Pi <
10$ mas) appear to rotate faster than the overall sample of
stars. However, a more accurate consideration reveals that for a
number of possible functional forms of $\Gamma(z)$ (e.g., $\Gamma
\cdot |z|$, $\Gamma \cdot |z|^2$), the most characteristic response is
expected in the $\vec{E}_3^{-1}$ harmonic, because the
$\vec{E}_1^{-1}$ harmonic is too sensitive to the choice of centroid
solar motion. Our choice of velocity in Table~\ref{vsun.tab},
consistent with the estimation for the more distant half of Hipparcos
stars, is justified by the fact that the OMM parameters are determined
in the distance-weighted (or proper motion) space where distant stars
are more significant, and whatever kinematics anomalies the nearest
stars may have, has little bearing on the OMM estimation problem. The
rotation gradient dipole $\vec{E}_1^{-1}$ emerges with a robust
positive coefficient $e_1^{-1}=11.25\pm 1.17$ \kmspc, which is
consistent with a negative gradient of $\Gamma(z)$.

\begin{figure}
\plotone{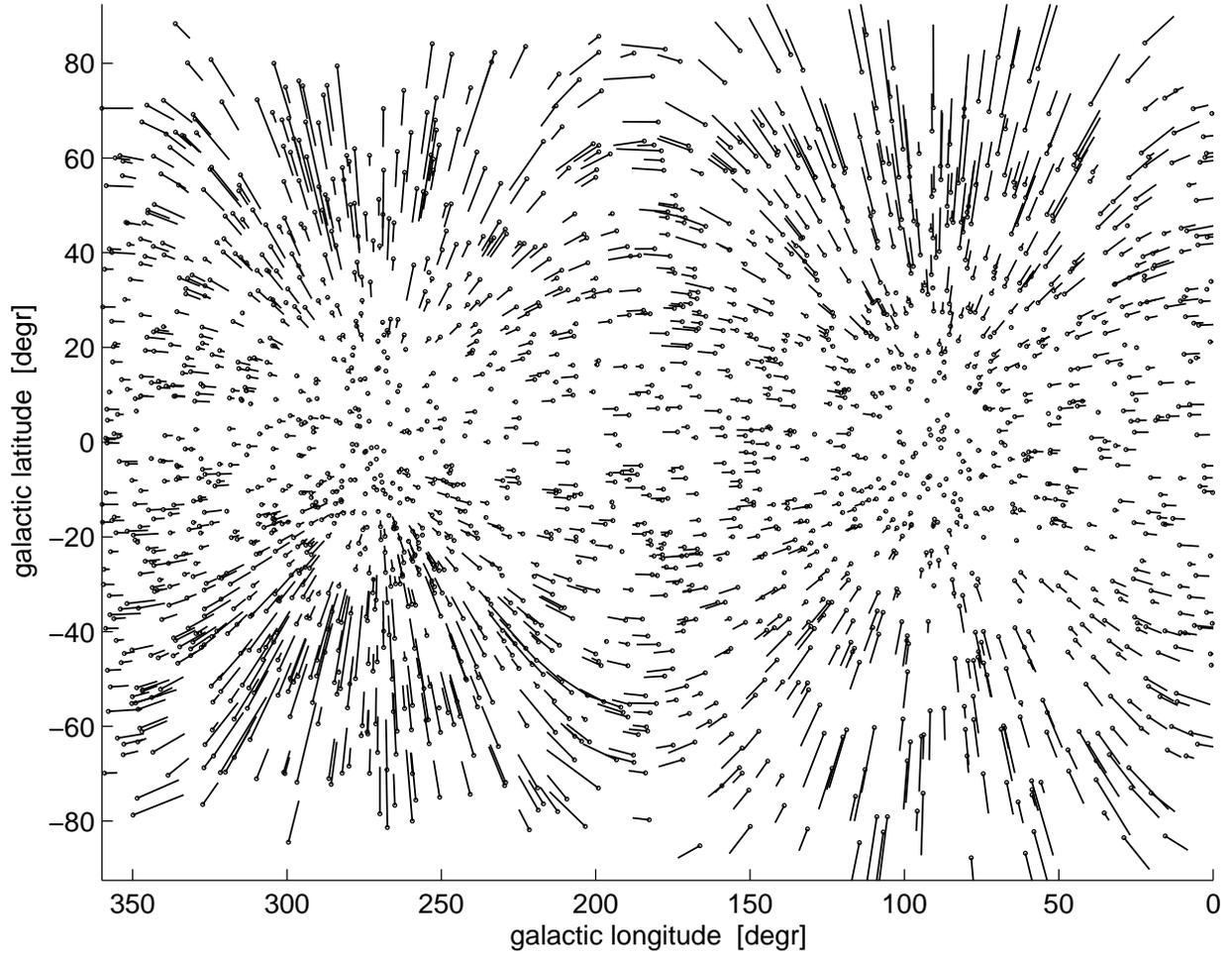}
\caption{The velocity field of Hipparcos stars generated by the vertical gradient of rotational velocity. \label{Gamma.fig}}
\end{figure}

We performed direct simulations of the vector harmonic response to a
linear gradient $\Gamma(z)=\Gamma \cdot |z|$ for different values of
$\Gamma$ and the height of the Sun above the plane $z_{\sun}$. A good
match with observations was found for $\Gamma=-20$ \kmspc and
$z_{\sun}=15$ pc, which yielded a set of coefficients
$e_1^{-1}=11.41\pm 1.17$, $h_2^{1}=-2.04\pm 0.47$, and
$e_3^{-1}=0.75\pm 0.22$ \kmspc, all the rest 21 harmonics being
insignificant. Both electric harmonics are in good agreement with our
fit for all star, whereas the $h_2^{1}$ coefficient is fairly close to
the fit ($-1.09 \pm 0.47$, not shown in
Table~\ref{om.tab}). Therefore, a linear gradient of rotational
velocity of the thin disk of roughly $-20$ \kmspc is a plausible
explanation to the corresponding set of vector harmonic terms beyond
the Ogorodnikov-Milne model. The detected pattern of tangential
velocities of Hipparcos stars consistent with this interpretation is
shown in Fig.~\ref{Gamma.fig}.

\section{Warp and the origin of $\vec{H}_1^{-1}$ and $\vec{H}_2^{-1}$}
The Milky Way disk is warped, as has been established from the
distribution of stars and neutral hydrogen. In this respect, our
Galaxy is not different from many other spiral galaxies exhibiting a
range of warp distortions. The origin of galactic warps is not clear;
a number of hypotheses have been proposed, including the tidal
interaction of the disk with the dark matter halo, the influence of
the bar, and the perturbation from a major satellite galaxy.  The Sun
appears to be close to the line of nodes of the Milky Way warp, and
the upper rim of the disk is at $\ell=90\degr$ (in the rotation
direction).  The height of the midsection above the plane is quadratic
with galactocentric distance in the model of \citet{dri},
$w(\rho)=(\rho-\rho_w)^2/15$ kpc for $\rho > \rho_w$, and zero for
$\rho < \rho_w$. According to \citet{mom}, the warp begins well within
the solar circle ($\rho_w < \rho_{\sun}$), and the line of nodes
deviates from the solar radius by $15\degr$.

\begin{figure}
\plotone{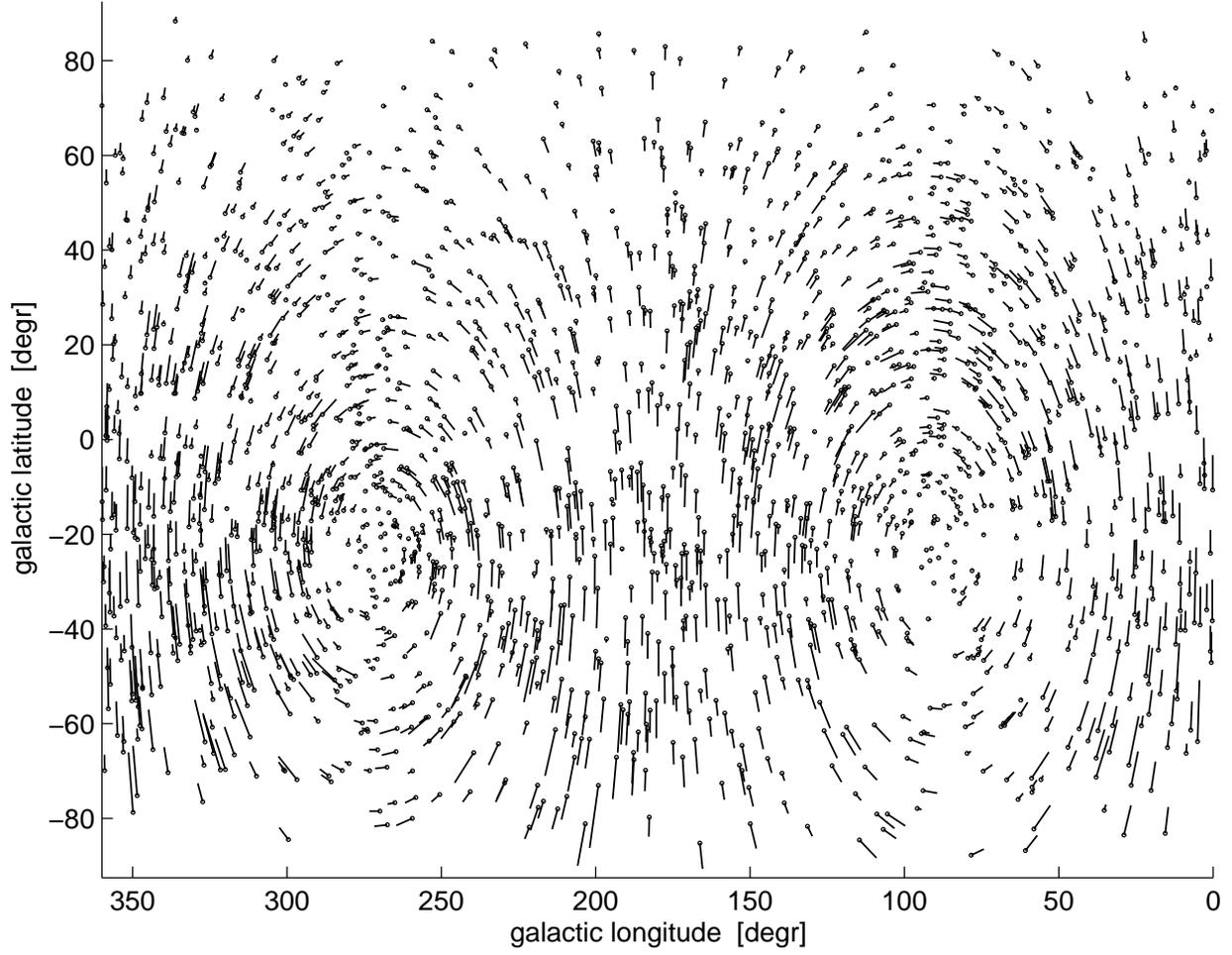}
\caption{The velocity field of Hipparcos stars generated by the two unexpected magnetic vector harmonics,
$\vec{H}_1^{-1}$ and $\vec{H}_2^{-1}$. \label{tt.fig}}
\end{figure}

The single most unexpected result of our analysis is the strong model
parameter $L_{13}$ (Table \ref{om.tab}), represented by the
coefficient of the first-degree magnetic harmonic $\vec{H}_1^{-1}$,
that is, a rigid rotation around the direction $-Y$ (see
Eqs.~\ref{mag.eq}).  The sign of this parameter implies that the stars
move upward in the direction of the Galactic center, downward in the
opposite direction, away from the center at the north pole, and toward
the center at the south pole. The signal-to-noise ratio on this
parameter is about 6. The extra statistically significant term
$h_2^{-1}$ detected by us in the local velocity field may be related
to the former. The pattern of tangential velocities generated by these
two magnetic harmonics, $6.21\,r\vec{H}_1^{-1}-1.20\,r\vec{H}_2^{-1}$,
is shown in Fig.~\ref{tt.fig}. The main effect of the higher degree
harmonic is that the axis of rotation lies below the plane at roughly
$b=-20\degr$, nearly obliterating the motion in the north pole region,
but retaining the galactocentric motion near the south pole. The most
conspicuous features are the general upward motion of stars in the
direction of the galactic center, and the downward motion of stars at
$\ell = 180\degr$. It is tempting to relate these two unexpected
components to a kinematic signature of the Galactic warp.

The shape of the warp, as traced by the distribution of neutral
hydrogen and dust, implies that the stars in the solar region are
involved in a general upward motion, since the starting rim of the
warp is within the solar circle ($\rho_w = 6.5 $ kpc). This common
motion is indistinguishable from the vertical solar reflex motion, but
the model also implies a radial gradient of the upward velocity,
$V_{Z,{\rm warp}} \propto (\rho-\rho_w)$, which is detectable in the
proper motion field. In the near-plane zone, the differential warp
motion manifests itself as a downward stream at $\ell \approx 0$, and
an upward stream at $\ell \approx 180\degr$. Obviously, the pattern in
Fig.~\ref{tt.fig} is completely inconsistent with this
prediction. Assuming for simplicity that the Sun lies on the line of
nodes, the linear Taylor expansion of the local velocity field
(\ref{taylor.eq}) should be expanded to include a vertical ($Z$)
component of velocity,

\eb
\lambda(\rho)=\lambda_0+\Lambda(\rho-\rho_0) \approx \lambda_0-\Lambda\,r\, \cos\ell \cos b.
\ee
The corresponding tangential velocity field is
\eb
\Delta\vec{v}_\tau = -\Lambda\, r\, \cos \ell \cos^2 b\, \tau_b=-\Lambda \,r \, (\frac{1}{2}\vec{H}_1^{-1}-\frac{1}{6}\vec{E}_2^1).
\ee

Thus, the differential warp motion is expressed by two OMM parameters,
$L_{13}$ and $M_{13}$. These two fitted parameters yield discrepant
estimates of the warp velocity gradient, $\Lambda = -2\, h_1^{-1}
\simeq -12$ \kmspc, and $\Lambda = 6\, e_2^{1} \simeq +4$ \kmspc. The
former estimate from the magnetic harmonic has the wrong sign, and its
modulus is too large for a credible differential warp. The electric
harmonic $\vec{E}_2^1$ has the right sign, but it nearly vanishes for
more distant stars ($\Pi < 10$ mas). Thus, the kinematical model of
Galactic warp does not furnish an adequate explanation to the presence
of magnetic harmonics $\vec{H}_1^{-1}$ and $\vec{H}_2^{-1}$. Samples
of larger volumes are needed to find out if these two harmonics are
not a local feature, and to find evidence of warp in the motion of
field stars. Interestingly, \citet{dri} also found a negative vertical
motion of distant OB stars in the direction of Galactic anticenter, in
obvious contradiction to the predicted warp motion. A non-stationary
warp is one of the possibilities considered by them. A precessing line
of nodes is conceivable, but we find it difficult to reconcile the
observed pattern of vertical motion, should it bear on the subject at
all, with a plausible precession model. It appears instead, that the
line of nodes is stationary, but the shape of warp changes to its
opposite every 50 Myr or so, curling this way and the other.

\section{Conclusions and Back to Astrometry}

Hipparcos stars with accurate trigonometric parallaxes represent only
a tiny fraction of the Galactic population. Half of stars considered
in this paper are within 112 pc, and 75 \% are within 160 pc. The
narrow horizon of our selection limits the accuracy of vector harmonic
terms describing the local tangential velocity field in the most
general and systematic fashion. Only several major kinematical
parameters can be determined with confidence from such a limited data
set.  We determine the relative solar velocity with respect to all
stars in our selection, and to stars with measured distances greater
than 100 pc. The latter determination yields
$V_{\sun}=(9.9,15.6,6.9)\pm 0.2$ \kms, which we add to the tangential
velocities of all field stars before performing a general
decomposition of the velocity field onto vector spherical
harmonics. This decomposition provides values of $A=13.8\pm 1.4$ and
$B=-13.4\pm 1.2$ \kmspc~ for the fundamental Oort's constants of
differential galactic rotation. Since $A+B\approx 0$, the local
rotation velocity curve is nearly flat. Assuming a galactocentric
distance of 7.5 kpc for the Sun, a rotation velocity of $\omega_0 =
204\cdot [\rho_0/7.5\;{\rm kpc}]$ \kms~ is derived.

Among other linear OMM parameters, we detect, most unexpectedly, a
strong signal carried by the first-degree magnetic harmonic
$\vec{H}_1^{-1}$, which describes a rigid rotation of the stellar
field around the axis $-Y$ opposite to the direction of galactic
rotation. The estimated rate of this rotation is roughly 6 \kmspc, or
1.3 \masyr~ in proper motions along the principal galactic
meridian. Another unexpected magnetic harmonic, $\vec{H}_2^{-1}$,
nearly cancels out the outward motion at the North pole predicated by
the former harmonic, but retains the strong inward motion around the
South pole, and the counter vertical motions near the Galactic
plane. Differential vertical velocities naturally arise from a
kinematic model of the Galactic warp, but we find that the sign of the
local rotation is opposite to what is required to raise the rim of the
Galaxy above the plane in the first and second quadrants. In other
words, the local stars are expected to move upwards due to the warp,
but we detect a negative differential rotation. Analysis of velocity
fields in a much larger volume of space is needed to make sure that
this discordant rotation is not a local feature, which would have
crucial consequences for our understanding of the physics of the warp.

Only three statistically significant vector harmonic terms beyond the
Ogorodnikov-Milne model are detected in this paper. One of them, the
electric multipole $\vec{E}_3^{-1}$, is of special note, since,
together with a positive residual dipole $\vec{E}_1^{-1}$ in the
direction of galactic rotation, it is likely to advertise a vertical
gradient of rotation velocity. A similar gradient of rotational lag of
$-30$ \kmspc~ has been found in proper motions of more distant stars
representing the thick disk population, but never reported for the
thin disk dominating our sample. We estimate a gradient of $\simeq
-20$ \kmspc~ for our sample of nearby stars limited to 200 pc. This
result requires verification on a larger sample of thin disk stars
extending to 1 kpc.

Our concluding remark is that estimation of subtle effects in the
local kinematics pertaining to the Galactic structure and formation
history is based on the assumption that the Hipparcos proper motion
data is free of large-scale systematic errors at $\ga 1$ \masyr. None
such errors have been reported in the literature, which is not a
strong argument because Hipparcos remains unparalleled at its level of
global astrometric accuracy. The major catalogs of proper motions
Tycho-2 \citep{ur} and UCAC \citep{za} are calibrated on Hipparcos
stars; therefore, systematic distortions of Hipparcos astrometry, if
any, are just copied over to these catalogs. Radio astrometric
observations with VLBI have recently advanced to a comparable level of
accuracy in positions and proper motions, and being directly tied to
the ICRF, provide an independent test for the Hipparcos reference
system \citep{bo}.  This important external check is unfortunately
limited by the small number of optically bright radio stars, but the
available accuracy of VLBI proper motions (approximately $1.7$ \masyr)
enables Boboltz et al. to state that the relative spin of the
Hipparcos proper motion system is much less than 1 \masyr~ about each
axis. This result confirms that the strong magnetic harmonic
$\vec{H}_1^{-1}$ representing a spin around the $-Y$ direction, is not
an artefact. Another significant astrometric development of late is
the SPM3 catalog, which provides high quality absolute proper motions
for a large sample of distant and faint stars, albeit in a small
fraction of the sky \citep{spm}. This catalog provides an independent
view of the local stellar velocity field in the surveyed area of the
sky.

\appendix

\section{Vector spherical harmonic decomposition of a proper motion
  field}

As customary in studies of Galactic dynamics, we make use of the
Galactic coordinate system $(X,Y,Z)$ in which the $X$ axis is pointing
toward the Galactic center, the $Y$ axis toward the direction of
Galactic rotation, and the $Z$ axis toward the north pole. For each
star, a triad of unit vectors $(\vec{r},\vec{\tau_l},\vec{\tau_b})$ is
defined, with

\eb
\vec{\tau}_r=\left(\begin{array}{c} \cos \ell \cos b \\ \sin \ell \cos b \\ \sin b \end{array}\right) \hspace{4mm}
\vec{\tau}_{\ell}=\left(\begin{array}{c} -\sin \ell \\ \cos \ell  \\ 0 \end{array}\right) \hspace{4mm}
\vec{\tau}_b=\left(\begin{array}{c} -\cos \ell \sin b \\ -\sin \ell \sin b \\ \cos b \end{array}\right),
\ee

where $\tau_\ell$ and $\tau_b$ define the tangential coordinate
directions toward increasing galactic longitude $\ell$ and the north
pole, respectively.  The proper motion vector of the object is
traditionally projected onto the locally tangential coordinate
vectors, that is, $\vec{\mu}=\mu_\ell\cos b\,\vec{\tau_\ell} +\mu_b\,
\vec{\tau_b}$.

A global proper motion field of a large set of celestial objects can
be represented by the expansion 

\eb
\vec{\mu}(\ell,b)=\sum_{n=1}^\infty \sum_{m=-n}^n [h_n^m \vec{H_n^m}(\ell,b)+e_n^m\vec{E_n^m}(\ell,b)],
\label{exp.eq}
\ee

where $\ell$ and $b$ are galactic longitudes and latitudes,
$\vec{H_n^m}$ and $\vec{E_n^m}$ are orthogonal vector harmonics which
we call magnetic and electric vector harmonics respectively. These
vector harmonics are derived via partial derivatives of the scalar
spherical harmonics over angular coordinates, viz.:

\begin{eqnarray}
\vec{H_n^m}(\ell,b)&=& \left[ \frac{\partial S_n^m(\ell,b)}{\partial b}\vec{\tau_\ell}-\frac{1}{\cos b}\frac{\partial S_n^m(\ell,b)}{\partial \ell}\vec{\tau_b}\right]\nonumber\\
\vec{E_n^m}(\ell,b)&=& \left[ \frac{1}{\cos b}\frac{\partial S_n^m(\ell,b)}{\partial \ell}\vec{\tau_\ell}+\frac{\partial S_n^m(\ell,b)}{\partial b}\vec{\tau_b}\right]
\end{eqnarray}
Spherical harmonics $S_n^m$ are counted by degrees $n=0,1,\ldots$ and orders $m=-n,-n+1,\ldots,n$. Explicitly,
\begin{eqnarray}
\label{sph.eq}
S_n^m &=& \sqrt{\frac{2n+1}{2\pi}\frac{(n-m)!}{(n+m)!}}     P_n^{m}(\sin b)\cos m\ell,    \hspace{8mm} m>0,\nonumber\\
      &=& \sqrt{\frac{2n+1}{4\pi}}                          P_n^{0}(\sin b)               \hspace{40mm} m=0,\nonumber\\
      &=& \sqrt{\frac{2n+1}{2\pi}\frac{(n-|m|)!}{(n+|m|)!}} P_n^{|m|}(\sin b)\sin |m|\ell, \hspace{4mm} m<0 \\
\end{eqnarray}
where $P_n^m$ are the associated Legendre polynomials. The first pair of
vector harmonics are generated by from the scalar zonal harmonic $S_1^0$, with the
electric component $\cos b \vec{\tau}_b$ and the magnetic component
$\cos b \vec{\tau}_\ell$. The electric vector harmonics for $n<=2$ in angular
coordinates are

\begin{eqnarray}
\label{ele.eq}
\vec{E}_1^{-1}&=& -\cos \ell \,\vec{\tau_\ell}+\sin \ell\sin b\,\vec{\tau_b}\nonumber \\
\vec{E}_1^{0}&=& \cos b\,\vec{\tau_b}\nonumber \\
\vec{E}_1^{1}&=& \sin \ell \,\vec{\tau_\ell}+\cos \ell\sin b\,\vec{\tau_b}\nonumber\\
\vec{E}_2^{-2}&=& 6\cos 2\ell \, \cos b \,\vec{\tau_\ell}-6\sin 2\ell\,\cos b\,\sin b\,\vec{\tau_b}\nonumber \\
\vec{E}_2^{-1}&=& -3\cos \ell \, \sin b \,\vec{\tau_\ell}-3\sin \ell\cos 2b\,\vec{\tau_b}\nonumber \\
\vec{E}_2^{0}&=& 3\cos b\,\sin b \,\vec{\tau_b}\nonumber \\
\vec{E}_2^{1}&=& 3\sin \ell \,\sin b \,\vec{\tau_\ell}-3\cos \ell\,\cos 2b\,\vec{\tau_b}\nonumber\\
\vec{E}_2^{2}&=&- 6\sin 2\ell \, \cos b \,\vec{\tau_\ell}-6\cos 2\ell\,\cos b\,\sin b\,\vec{\tau_b}, 
\end{eqnarray}
and the magnetic vector harmonics  for $n<=2$ are
\begin{eqnarray}
\label{mag.eq}
\vec{H}_1^{-1}&=& \sin \ell \,\sin b \,\vec{\tau_\ell}+\cos \ell\,\vec{\tau_b}\nonumber \\
\vec{H}_1^{0}&=& \cos b\,\vec{\tau_\ell}\nonumber \\
\vec{H}_1^{1}&=& \cos \ell \,\sin b\,\vec{\tau_\ell}-\sin \ell\,\vec{\tau_b}\nonumber\\
\vec{H}_2^{-2}&=& -6\sin 2\ell \, \cos b\,\sin b \,\vec{\tau_\ell}-6\cos 2\ell\,\cos b\,\vec{\tau_b}\nonumber \\
\vec{H}_2^{-1}&=& -3\sin \ell \, \cos 2b \,\vec{\tau_\ell}+3\cos \ell\sin b\,\vec{\tau_b}\nonumber \\
\vec{H}_2^{0}&=& 3\cos b\,\sin b \,\vec{\tau_\ell}\nonumber \\
\vec{H}_2^{1}&=& -3\cos \ell \,\cos 2b \,\vec{\tau_\ell}-3\sin \ell\,\sin b\,\vec{\tau_b}\nonumber\\
\vec{H}_2^{2}&=&- 6\cos 2\ell \, \cos b \,\sin b\,\vec{\tau_\ell}+6\sin 2\ell\,\cos b\,\vec{\tau_b}, 
\end{eqnarray}

Note that the normalization coefficients of the generic spherical harmonics (\ref{sph.eq}) are omitted in these formulas.

\section{Ogorodnikov-Milne model via vector harmonics formalism}

The linear Ogorodnikov-Milne model of the local velocity field in its
general matrix form can be written as \citep{dum}

\eb
\vec{V}=\left(\begin{array}{ccc}
M_{11} & M_{12} & M_{13}\\
M_{12} & M_{22} & M_{23}\\
M_{13} & M_{23} & M_{33} \end{array} \right) \vec{r} + \left( \begin{array}{ccc}
0 & -L_{12} & -L_{13}\\
L_{12} & 0 & -L_{23}\\
L_{13} & L_{23} & 0 \end{array} \right) \vec{r},
\ee
where $\vec{V}$ is the systemic part of the velocity field as a
function of the position vector $\vec{r}$. It is assumed in the
following that positions are determined with respect to the solar
system barycenter; any shift of the coordinate system origin results
in an additional constant translation term. For convenience, the
matrix of transformation is split into the symmetric (shear) part $M$
and the antisymmetric traceless (rotation) part $L$. It is readily
seen that the matrix $M$ describes the gradient-type distortions of
the field, and the $L$ part represents rigid rotations, or spins,
around the three coordinate axes.

After a small manipulation, the tangential velocity components are
\begin{eqnarray}
\label{v.eq}
V_l /r =  (\vec{V}\cdot \vec{\tau}_\ell)/r & = & (-M_{11}+M_{22})\sin \ell \cos \ell \cos b + M_{12}\cos 2\ell \cos b \nonumber\\ & &-M_{13}\sin \ell \sin b
	+M_{23}\cos \ell \sin b  + L_{12} \cos b +L_{13}\sin \ell \sin b -L_{23}\cos \ell \sin b \nonumber\\
V_b /r =  (\vec{V}\cdot \vec{\tau}_b)/r & = & (-M_{11}\cos^2 \ell-M_{22}\sin^2 \ell+M_{33})\sin b \cos b - M_{12}\sin 2\ell \sin b\cos b
	\nonumber\\ & &+M_{13}\cos \ell \cos 2b +M_{23}\sin \ell \cos 2b  +L_{13}\cos \ell +L_{23}\sin \ell \nonumber
\end{eqnarray}
Comparing these equations with the trigonometric expressions for low-degree vector spherical harmonics (Appendix A), the following
relations of proportionality are established
\begin{eqnarray}
M_{12} & \propto & e_2^{-2} \nonumber\\
M_{23} & \propto & -e_2^{-1} \nonumber\\
M_{13} & \propto & -e_2^{1} \nonumber\\
D_{1} & \propto & e_2^{2} \nonumber\\
D_{2} & \propto & e_2^{0} \nonumber\\
L_{12} & \propto & h_1^{0} \nonumber\\
L_{13} & \propto & h_1^{-1} \nonumber\\
L_{23} & \propto & -h_1^{1},
\end{eqnarray}
where 
\eb
D_1=\frac{1}{2}(M_{11}-M_{22}); \hspace{4mm} D_2=M_{33}+\frac{1}{2}(M_{11}+M_{22}).
\ee

Clearly, all nine parameters of the Ogorodnikov-Milne model can not be
determined from a proper motion field, since an isotropic dilation
($M_{11}=M_{22}=M_{33}$) results in radial velocities only. This is
why only eight parameters of the model appear in the vector harmonic
decomposition of a proper motion field \citep{vit}.

We further establish the relations between the constants of the
Ogorodnikov-Milne model and the four constants ($A,B,C,K$) of the
Oort's two-dimensional model by matching the terms in the proper
motion equation of the latter \citep{tor}

\begin{eqnarray}
4.741\,\vec{\mu} &=& (A\cos 2\ell \cos b +B\cos b -C\sin 2\ell \cos b)\,\vec{\tau}_\ell \nonumber\\
&& +(-A\sin 2l\ell \sin b \cos b -C\cos 2\ell \sin b \cos b -K\sin b \cos b)\,\vec{\tau}_b.
\end{eqnarray}
The corresponding equation for the more general Ogorodnikov-Milne model in terms of vector harmonics is
\begin{eqnarray}
\label{om.eq}
4.741\,\vec{\mu} &=& L_{13}\vec{H}_1^{-1}+L_{12}\vec{H}_1^0-L_{23}\vec{H}_1^1 \nonumber \\
&& +\frac{1}{6}M_{12}\vec{E}_2^{-2}-\frac{1}{3}M_{23}\vec{E}_2^{-1}-\frac{1}{3}D_{2}\vec{E}_2^{0}-\frac{1}{3}M_{13}\vec{E}_2^{1}
+\frac{1}{6}D_{1}\vec{E}_2^{2},
\end{eqnarray}
where $M_{12}=A$, $L_{12}=B$, $D_1=C$ and $D_2=K$.

\acknowledgments The research described in this paper was carried out
at the Jet Propulsion Laboratory, California Institute of Technology,
under a contract with the National Aeronautics and Space
Administration.

\clearpage
\begin{deluxetable}{lrr}
\tabletypesize{\scriptsize}
\tablecaption{Determination of the centroid velocity of the Sun \label{vsun.tab}}
\tablewidth{0pt}
\tablehead{
\colhead{component} &  \colhead{$\Pi < 10$ mas}  & \colhead{all stars} }
\startdata
$V_X = e_1^{1}$ & $9.9\pm 0.2$ & $ 10.5 \pm 0.1$  \\
$V_Y = e_1^{-1}$ & $15.6 \pm 0.2$ & $ 18.5 \pm 0.1$  \\
$V_Z = -e_1^{0}$ & $6.9 \pm 0.2$ &  $7.3 \pm 0.1$ \\
\enddata
\end{deluxetable}

\clearpage
\begin{deluxetable}{lrr}
\tabletypesize{\scriptsize}
\tablecaption{Determination of Ogorodnikov-Milne and higher degree parameters of the local velocity field \label{om.tab}}
\tablewidth{0pt}
\tablehead{
\colhead{} &  \colhead{$\Pi < 10$ mas}  & \colhead{all stars} }
\startdata
$L_{13} = h_1^{-1}$ & $5.91 \pm 1.02\;(5.8)$ & $ 6.21 \pm 0.94\; (6.6)$  \\
$L_{12} = B= h_1^{0}$ & $-12.36 \pm 1.26\;(9.8)$ & $ -13.36 \pm 1.16\; (11.5)$  \\
$L_{23} = -h_1^{1}$ & $0.13 \pm 0.97\;(0.1)$ & $ -0.36 \pm 0.89\; (0.4)$  \\
$M_{12} = A= 6\,e_2^{-2}$ & $14.08 \pm 1.56\;(9.2)$ & $ 13.83 \pm 1.42\; (9.8)$  \\
$M_{23} = -3\,e_2^{-1}$ & $0.38 \pm 1.37\;(0.3)$ & $ 0.76 \pm 1.25\; (0.6)$  \\
$X_{2} = K= -3\,e_2^{0}$ & $-0.63 \pm 1.98\;(0.3)$ & $ 1.02 \pm 1.81\; (0.6)$  \\
$M_{13} = -3\,e_2^{1}$ & $-0.32 \pm 1.38\;(0.2)$ & $ -2.13 \pm 1.27\; (1.7)$  \\
$X_{1} = C= 6\,e_2^{2}$ & $-4.72 \pm 1.64\;(3.1)$ & $ -3.03 \pm 1.43\; (2.1)$  \\
\multicolumn{3}{l}{Other significant parameters \dotfill} \\
$e_1^{-1}$ & $-0.81 \pm 1.28\;(0.6)$ & $ 11.25 \pm 1.17\; (9.6)$  \\
$h_2^{-1}$ & $-1.69 \pm 0.44\;(3.8)$ & $ -1.20 \pm 0.41\; (2.9)$  \\
$e_3^{-1}$ & $0.52 \pm 0.24\;(2.2)$ & $ 0.80 \pm 0.22\; (3.7)$  \\
$e_4^{2}$ & $-0.075 \pm 0.041\;(1.9)$ & $ -0.100 \pm 0.037\; (2.7)$  \\
\enddata
\tablecomments{All parameters and their formal standard errors are specified in \kmspc; the signal-to-noise ratio is
given in brackets. A solar velocity $V_{\sun}=(9.9, 15.6, 6.9)$ \kms~ was subtracted for both sets of stars.}
\end{deluxetable}

\end{document}